\documentclass[12pt]{article}

\usepackage[left=1.5cm, right=1.5cm, top=1.5cm, bottom=1.5cm]{geometry}
\usepackage{graphicx}
\usepackage{multicol}
\usepackage[labelfont={bf,sl}]{caption}
\usepackage{subfigure}
\usepackage{lscape}
\usepackage{amssymb}
\usepackage{subfloat}
\usepackage{float}

\newcommand{\keywards}{\textbf{Keywords: }}
\newcommand{\pacs}{\textbf{PACS: }}
\Large
\title{Search for Missing Links Between Two Extreme Wind Speed Profiles : Dark Energy Accretion and Adiabatic Fluid Accretion}

\date{}

\author{Parthajit Roy $^1$ $\;\;\;\;\;\;\;\;\;\;\;\;\;\;\;\;$ Ritabrata Biswas $^{2*}$\\ \\$^1$Department of Computer Science, The University of Burdwan\\City:Burdwan-713104,\\Dist: Purba Bardhaman, State: West Bengal, India\\
	e-mail: \texttt{roy.parthajit@gmail.com}\\ \\$^2$Department of Mathematics, The University of Burdwan\\Golapbag Academic Complex, City:Burdwan-713104,\\Dist: Purba Bardhaman, State: West Bengal, India\\e-mail: \texttt{biswas.ritabrata@gmail.com} \footnote{$^2$ \emph{ Corresponding Author:} biswas.ritabrata@gmail.com}}

\begin{document}
	
	\maketitle

\begin{abstract}
	In recent past, the progresses in  accretion studies onto relativistically gravitating central objects like a Schwarzschild singularity reveal that the accretion flow must be transonic. For such cases, the radial inward speed gradient can be written as a numerator by denominator form among which the later vanishes somewhere in between infinite distance from the attracter to the event horizon of the same. For sustainability of a physical solutions, the numerator should vanish at the same radial distance where denominator does vanish. From this point, we   obtain a second degree first order differential equation of radial inward speed and hence we obtain two branches of flow, namely accretion and wind. For adiabatic accretion case, the wind curve is formed to be more or less parallel to the radial distance axis as we move far from the central object. For dark energy accretion, this curve is parallel to the radial velocity axis. Here we face a question. Why there is no fluid speed profiles in between these two extremities. While searching the reasons, we follow that dark energy, if treated as an accreting object,  should stay around the central \emph{compact star} and hence will contaminate the metric which propertises the compact star. In this research work, we have proposed a model with a rotating black hole embedded in quintessence where quintessence equation of state and spin parameters of the black hole are regulatory factors of the model. The resulting accretion and wind curves are studied. The Effect of negative pressure of dark energy is found to get catalyzed by the entry of the spin of the black hole. We tally our results with observations of accretion or outflow phenomenon near to different quasars.
\end{abstract}

\keywards{Accretion Disc, Supermassive Black Holes, Adiabatic Gas, Dark Energy, Quintessence, Modified Chaplygin Gas.}


\section{Introduction}
\label{sec:introduction}

Though general relativity (GR hereafter) predicts about the existence of space time wrapped by event horizon, popularly coined as a black hole (BH hereafter) and properties of accretion discs around such compact objects are studied since 1972~\cite{Michel:1972}, exact observational evidences were hardly found. The scenario  has been changed very recently in 2919 after the first time ever ``image of an event horizon'' was seen\cite{apj:2019:1, apj:2019:2} (Actually, the first image of a BH's shadow was captured by the event horizon telescope).


This extreme supermassive object is supposed to be formed out of merger of two SMBHs~\cite{apj:2019:5}~\cite{gebhardt:2011:6}. A prominent jet outflow from the central region is found and a rotating disc of ionized gas around this SMBH is observed,  which is aligned perpendicular to the direction of the said jet. The angular speed of the disc is found to be up to $10^3{km\times sec^{-1}}$~\cite{macchetto:1997:7}. This disc spans a maximum diameter of $2.5 \times 10^4AU$~\cite{matveyenko:2011:8} which is around 6000 times of Pluto's average orbital radius around the Sun (Schwarzschild radius of this particular SMBH is estimated as $120AU$~\cite{akiyama:2015:10}). Accretion rate is found to be of $0.1M_\odot yr^{-1}$~\cite{di:Matteo:2003:9}.


Astrophysically,   BHs can be signatured by their masses, sizes and accretion disc properties. They are essentially the remnants  of post main sequence evolution of massive stars. If we wish to analyze them general relativistically, we will define a region wrapped by a singular 4-surface  (called event horizon), which does not let us to see the inside phenomena. Even space and time swap their roles if we cross (at least theoretically) this horizon. The central singularity $r=0$ resides inside the said wrapped region. When a BH is in a  binary or in the center of a galaxy surrounded by materials nearby it, due to the Roche-lobe overflow, fluids from the companion star starts to flow towards this compact object affected by its gravitational pull and accretion is said to take place. Since late 1940's, Hoyle and  Lyttleton~\cite{hoyle:1940} had started the theoretical studies of accretion onto gravitating objects. Through different time era, these studies were enriched by Bondi~\cite{bondi:1952}, Michel~\cite{Michel:1972}, Shakura and Sunyaev~\cite{Shakura:1974} and Chakrabarti~\cite{chakraborty:1996}. In the process, different authors have collected the quantitative data regarding different characteristics of an accretion disc near SMBHs and that of behaviors of in-falling matters etc via the radiations coming out of them. These radiations are the thermal energy which are the converted form of a part of released potential energy from the accreting fluid falling into the potential well.


Accretion around a Schwarzschild type of singularity was studied by Michel~\cite{Michel:1972} for the first time where he has obtained the integration of general relativistic continuity and energy flux equations and combined them to form a radial inward velocity gradient, i.e., an ordinary differential equation consisting  $\frac{dv}{dr}$, where $v$ is the radial  inward velocity and $r$ is the radial distance from the BH. One needs to choose a continuous flow which requires the denominator and numerator of $\frac{dv}{dr}$ should simultaneously vanish (such that L'Hospital's rule can be applied).

To study the accretion properties, we require to point out the amount of force, that is applied on a particle by the central gravitating object, i.e., the concerned BH. Simply this will be different than a Newtonian case. General relativistic equations turn highly non-linear due to the presence of time and radial distance together as independent variables. To obtain a stationary result, we should impose a pseudo Newtonian (PNF) force in the corresponding Navier-Stokes' equations.

In 2002, Mukhopadhyay calculated a PNF~\cite{Mukhopadhyay:2002:BM2002} which depends on the radial distance from the BH and the spin parameter of the BH. A picture of accretion onto a rotating BH with the help of this PNF has also been constructed in the reference~\cite{Mukhopadhyay:2003:BM2003}. This second work considered the non-viscous adiabatic accretion flow through the critical point.

While considering a BH or the corresponding metric, we should incorporate all known/assumed properties of the BH along with the background space time's effects on it. Generally, a BH is assumed to reside in general relativistic space-time. However, the background may be changed by incorporating modified gravity or a modification  of energy momentum tensor $T_{\mu \nu}$. Modification of $T_{\mu \nu}$ may be established by the incorporation of exotic energy. The next paragraph explains what do we exactly mean by inclusion of exotic matter/energy.

To measure the cosmological distances,  some particular standard candles are popularly used. For short  intra-galactic distances or distances of nearby galaxies, we use Cepheid variables. When it turns to measure the distances of far away galaxies, we use type Ia supernovae  (SNEIa). Two distinguished supernova data analyzing collaborative groups published their observations in late 1990's~\cite{riess:1998:11}~\cite{Perlmutter:1999:12} which informed us that our universe is going through a phase of late time cosmic accretion. 

In 1999, Huterer, Turner et al~\cite{huterer:13} came up with a new concept called Dark Energy (DE). They included DE in the stress energy part of Einstein's field equation. DE occupies nearly $70\%$ of the energy density budget of the present time cosmos. There is a topological defect~\cite{PhysRevLett:14} and a scalar field~\cite{bronstein:1933:18} which is called quintessence~\cite{PhysRevLett:caldwell:1998:24}, can explain this fact. 

Quintessence is canonical scalar field explained with gravity, free from theoretical  problems like Lagrange's instabilities and appearance of ghost. Core idea of quintessence is to describe a slowly evolving scalar field $Q$ along with a potential $V(Q)$ which can justify  the accelerated expansion of universe with the Equation of State (EoS) given by $p_q = \omega_q \rho_q; (-1<\omega_q < 1/3),$ $p_q$ and $\rho_q$ are respectively quintessence pressure and energy density. The EoS parameter can be expressed as equation~(\ref{eqn:eos}).

\begin{equation}
	\omega_q = \frac{p_q}{\rho_q} = \frac{\frac{1}{2} \dot{Q}^2 + V(Q)}{\frac{1}{2} \dot{Q}^2 - V(Q)}\;\;,
	\label{eqn:eos}
\end{equation}  

where $\dot{Q}$ represents $dQ/dE$, $E$ being the cosmic time. Observations constrain $\omega_q$ as $\omega_q < -0.51$ according to~\cite{sereno:2002:40} and $\omega_q < -1$ according to Mortsell et. al~\cite{mortsell:2001:41}. There are two kinds of quintessence models. The first one is the thawing model where Hubble friction nearly freezes the field during early cosmological epoch. The second model is freezing model where potential tends to be flimsy at late times in this model. This causes a monotonic retardation of fields. 

In this article, we have considered another DE model known as the Modified Chaplygin Gas(MCG). It has its origin from the brane world scenario~\cite{subenoy:2009:mcg1} and its EoS is given as:

\begin{equation}
	p_{MCG} = \alpha \rho_{MCG} - \frac{\beta}{\rho_{MCG}^n}.
\end{equation}

This  fluid is constituted like a two fluid model. The first part of this model obeys the perfect fluid type EoS $P=\alpha \rho$ and the later follows the generalized Chaplygin  gas  type behavior~\cite{benaoum:2002:mcg2}. $\alpha,\beta$ and $n$ are parameters of the model and $p_{MCG}$ and $\rho_{MCG}$ are the pressure and energy density of MCG fluid. In the reference~\cite{lu:2011:mcg3}, the authors have shown, if flat universe is the background, MCG model's constraints are 

\[
    \beta =    0.00189_{-0.00756}^{+0.00583} (1\sigma)_{-0.00915}^{+0.00660}(2\sigma) \;\;and\;\;  n = 0.1079_{-0.2539}^{+0.3397}(1\sigma)_{-0.29111}^{+0.4678}(2\sigma).
\]

In the background of quintessence universe, a rotating BH solution was proposed for the first time by Ghosh~\cite{Ghosh2016:ref10} in 2016 as,

\begin{equation}
ds^2=-\frac{\Delta-a^2sin^2\theta}{\Sigma}dt^2+\frac{\Sigma}{\Delta}dr^2-2asin^2\theta\left(1-\frac{\Delta-a^2sin^2\theta}{\Sigma}\right)dt d\phi
\label{eqn:ds:2}
\end{equation}
$$+\Sigma d\theta^2 +sin^2 \theta\left[\Sigma+a^2sin^2\theta\left(2-\frac{\Delta-a^2sin^2\theta}{\Sigma}\right)\right]d\phi^2$$
where $\Delta=r^2+a^2-2Mr-\frac{{\cal A}_q}{\Sigma^{\frac{(3\omega_q-1)}{2}}}$ and $\Sigma=r^2+a^2cos^2\theta$.

In this solution , if ${\cal A}_q$ is chosen to vanish, we revert back to Kerr solution. The article [28] itself has  analysed the geometry of quintessence embedded black holes very well. A scalar polynomial singularity at $r=0$ is found for $\omega_q \neq \{0,~\frac{1}{3},~-1\}$. A general form of exact spherically symmetric solutions for Einstein's field equations which describes BHs surrounded by quintessential matter is observed. The parameter $\omega_q$ should possess the range $-1<\omega_q<-\frac{1}{3}$ for a deSitter horizon. Similarly, the range  $-\frac{1}{3}<\omega_q<0$ supports asymptotically flat solution. $\omega_q=-1$ is a borderline case of extraordinary quintessence to cover the cosmological constant term $a=J/M$ is spin parameter of the BH. Point to be noted that the Kerr-deSitter solution is not found by $\omega_q=-1$. As quintessential field is present, the geometry is not Ricci flat \cite{toshmatov:2017, adbujabbarov:2017}. If $A_q=0$, the solution turns into Kerr BH. Articles like~\cite{toshmatov:2017}  shows that for $\omega_q \approx -1$, the asymptotically relevant properties for a circular geodesic of the quintessential rotating BH spacetimes are very close to those which are occuring in the Kerr-DeSitter spacetimes\cite{Stuchl:2004}. We can conclude that asymptotical phenomena treated in vacuum spacetime can approximately reflect these phenomenon around BHs in quintessential field with $\omega_q \approx = -1 $.

Sarkar and Biswas have calculated a PNF for this BH solution~(\ref{eqn:ds:2}) in their work~\cite{refId0:sss1} as,

$$F_x=\left[ a {\cal A}_q  x^{3 \omega_q} \left\{ a^2 (3 \omega_q+1)+ 3x^2 (\omega_q+1)-8x\right\}+2 a \left\{a^2+x (3 x-4)\right\} x^{6 \omega_q}-2 a {\cal A}_q ^2 x\right.$$
\begin{equation}
\frac{\left. -\sqrt{2} x^{6 \omega_q+6}  \left\{ \left(a^2+x^2-2x\right) x^{3 \omega_q}-{\cal A}_q x\right\}\sqrt{x^{-9 (\omega_q+1)} \left({\cal A}_q +3 {\cal A}_q  \omega_q+2 x^{3 \omega_q}\right)}\right]^2}{x^3 \left[a^2 x^{3 \omega_q} \left({\cal A}_q +3 {\cal A}_q  \omega_q +2 x^{3 \omega_q}\right)-2 x \left\{{\cal A}_q +(2-x) x^{3 \omega_q}\right\}^2\right]^2}.
\label{eqn:pnp:quint}
\end{equation}

Clearly, this PNF depends on $x$, the distance from BH taken into $\frac{GM}{c^2}$ unit;  on $a$, the spin parameter in $c$ (speed of light) unit and on the quintessence parameters $\mathcal{A}_q$ and $\omega_q$. The procedure of the PNF's construction for quintessence embedded Kerr BH keeps the $r^{\omega_q}$ term in nonlinear orders. So mathematically it is difficult to go back to the Kerr case as a terminal result. But we observe that the PNF is able to stay in a least deviation range as we take ${\cal A}_q=0$. Besides, the accretion result does exactly match with the terminal cases.

In this present article, we will try to study the natures and properties of non-viscous accretion onto a rotating BH which is embedded in DE universe. In literature, we  find several works which analyze the detailed properties of accretions of  DE. In 2004, Babichev et al~\cite{babichev:PhysRevLett:2004} have shown that the mass of BH may reduce due to the DE accretion process. However, this does not lead to form any naked singularity as predicted by the reference~\cite{PRD:gao:2008}. Biswas et al. have assumed the MCG type DE accretion onto  SMBHs~\cite{RBcqg:Biswas_2011}. They have considered the accreting matter to be dominated by DE in the scenario of late time cosmic acceleration. SMBHs are formed to reside in the central regions of the galaxies. It is widely believed that cold dark matter's (DM) hierarchical clustering is the supreme origin       of the structure of universe~\cite{45rb:blumenthal:1984}. DM halo's angular momentum and the concerned galaxy's rotation are theoretised to be produced by gravitational tidal torque~\cite{46rb:barnes:1987}. Explanations say that the halos obtain their spins through the cumulative acquisition of angular momentum from satellite accretion. DE also does accrete on a galaxy and hence it gets similarity torqued by such tidal interactions. It is very much reasonable to expect that an amount of angular momentum should  reside inside DE halos of galaxies. Whenever such a rotating DE is to accrete on a compact object, it would carry a part of its own angular momentum with it which ultimately leads to the formation of a DE accretion disc. In the reference~\cite{RBcqg:Biswas_2011}, this was the main logic to consider a DE accretion disc. It was shown that the  wind is stronger for such accretion and at a finite distance from the BH, the wind speed becomes equal to that of light. This signifies that the non-viscous accretion of DE weakens the accretion procedure.  Biswas~\cite{Biswas:2011:EPL} has shown the density profiles of a DE accretion should have a moderately differentiated minima.

Dutta and Biswas have considered viscous MCG accretion in the reference~\cite{Dutta2019:Dutta}. Presently, the constraint of dissipation from the gravitational wave-event GW150914 predicted a nonzero value of the shear viscosity of DE through which the gravitational wave is propagating~\cite{PhysRevD:goswami:36OfDutta}. The upper-bound of the shear viscosity is found to be $n < 5.2n_{crit} ~ 2.3 \times 10^9 Pasec$ where $n_{crit} = \rho_{crit} H_0^{-1} = 3.21 \times 10^{-5}(GeV)^3 = 4.38 \times 10^8 Pasec$, using $\rho_{crit} =  H_0^2M_{PL}^2$. Bulk viscosity takes larger value than shear viscosity. These ideas lead the authors  to study the viscous DE accretion and the lower limit of the shear viscosity to entropy density ratio~\cite{Dutta2017:SDRBEPJC2}. All these works have chosen the central compact object to be a Kerr BH. However, the natures of the wind speed gradient for dark energy accretion and adiabatic accretion are far distantly aligned. Either we see the wind to have constant speed or it is flowing with a speed equal to that of light at a finite distance from the central BH. Between these two extremities, no intermediate alignment is found. We are motivated here to find the background conditions for accretion which may  lead to prove intermediate curves. To find the missing alignment, we consider accretion onto DE contaminated BH. Present article will discuss the natures of accretion onto a BH which itself is residing into a DE universe and has the effect of the exotic matter inside its metric. Considering so, we can even analyze  the accretion results where adiabatic fluid is accreting but an effect of DE is clearer as we choose mare and mare negative EoS ($\omega_q$) value of quintessence. A question may arise here that how much it will be possible for some exotic matter/energy to contaminate the strong gravitational attraction of a BH. DM, in the form of bosons, is able to form self generating    bond structures in galaxies~\cite{physrevlett:2018:dm1}. Rotational curves of at least Milky way have been justified by considering presence of huge amount of DM in core area of the galaxy~\cite{boshkayev:2019:md3}. If the motions of test particles are studied in the gravitational field of both SMBH and DM core, a significant discrepancy in the motion is found only below 100AU which keeps on increasing as we go towards the center. This means that the current observations of the motion of the stars, e.g. $S_2$, which exists in the near vicinity of the $SGrA*$~\cite{refId0:gal} cannot distinguish a DM core model from the SMBH. Now DE and DM interact with each other~\cite{sym10090411:inter}. Again, the DM clustering is likely to get affected by the presence of DE. This indirectly is a controller of galactic structure formation and as a result, the DM will obviously affect the central SMBH and the accreting, in-falling matters which are following the galactic rotational speed distribution (which is again caused by the presence of DM).

We organize this article as following: first, we recapitulate the basic mathematical structures. Then we will find out the inward velocity solution curves and analyze them. The same will be done for the sound speed curves. Solutions will be derived for both adiabatic and DE (of MCG type) accretions. Finally, we will briefly discuss and conclude our results.

\section{Mathematical Construction of the Model}
\label{sec:basic:math}

In this section, we will construct the mathematical model from references~\cite{Mukhopadhyay:2003:BM2003} and~\cite{Biswas:2011:RB2010}. First we will consider the continuity equation,

\[
\frac{\partial \rho}{\partial z} + \vec{\nabla}. (\rho \vec{V}) = 0
\]
which is simplified to
\begin{equation}
    \frac{d}{dx} \left(x u \Sigma\right) = 0~,
\label{eqn:math1}
\end{equation}
for stationary and cylindrical structure, where $\Sigma$ is vertically integrated density expressed as 

\begin{equation}
	\Sigma = I_C \rho_e h(x),
\label{eqn:math2}
\end{equation}
 with $I_C$ = constant (related to EoS of accreting fluid) = 1 (for simplicity), $\rho_e$ = density of the accreting fluid at the equatorial plane, $h(x)$ = half, thickness of the disc. $u= u_x = \frac{v_x}{c}$, $v_x$ is the radially inward speed of accretion. Next, we will consider the radial component of stationary Navier Stoke's equation.

\begin{equation}
	\rho(\vec{V}.\vec{\nabla})\vec{V} = \vec{\nabla}\rho + \rho \gamma \nabla^2 u - F_{GX}
\end{equation} 
as,

\begin{equation}
	u\frac{du}{dx} + \frac{1}{\rho} \frac{dp}{dx}-\frac{\lambda^2}{x^3}+F_g\left( x\right)=0
\label{eqn:math3}
\end{equation}
where $F_g(x)$ is the radially inward gravitational force component. Assuming the vertical equilibrium from the vertical component, we get,

\begin{equation}
	h(x) = c_s \sqrt{\frac{x}{F_g}}.
\label{eqn:math4}
\end{equation}

With the help of the equations~(\ref{eqn:math1}), (\ref{eqn:math2}), (\ref{eqn:math3}) \& (\ref{eqn:math4}), we can build the ordinary differential equations for sound speed and radial velocity as,

\begin{equation}
\frac{d c_s}{dx} = \left( \frac{3}{2x} -\frac{1}{2F_g} \frac{dF_g}{dx} + \frac{1}{u} \frac{du}{dx} \right) \left\lbrace \frac{\left( n+1 \right) c_s \left( c_s^2 -\alpha \right)}{\left( 1-n \right) c_s^2 + \alpha \left( n+1 \right)} \right\rbrace \;\; and
\label{eqn:diffeqn:sound}
\end{equation}

\begin{equation}
\frac{du}{dx} = \frac{\frac{\lambda^2}{x^3}- F_g\left( x \right) + \left( \frac{3}{x} - \frac{1}{F_g} \frac{dF_g}{dx} \right) \frac{ c_s^4}{\lbrace \left( 1-n \right) c_s^2 + \alpha \left( n+1 \right)\rbrace}}{u - \frac{2 c_s^4}{u \lbrace \left( 1-n \right) c_s^2 + \alpha \left( n+1 \right)\rbrace}} \;\;\;\; respectively.
\label{eqn:diffeqn:fluid}
\end{equation}

Now, we must consider the flow to be continuous. To satisfy this, the numerator and the denominator of equation~(\ref{eqn:diffeqn:fluid})  must vanish together at some point $x=x_c$. This will allow us to apply L'Hospital's rule on the equation (\ref{eqn:diffeqn:fluid}) and will provide us the quadratic equation for radial speed gradient as,

\begin{equation}\label{accretioncqg11}
\mathcal{A}\left(\frac{du}{dx}\right)^{2}_{x=x_c}+\mathcal{B}\left(\frac{du}{dx}\right)_{x=x_c}+\mathcal{C}=0,
\end{equation}
where $$\mathcal{A}=2\left[1-\frac{2\left(c_{sc}^{2}-\alpha\right)\left(n+1\right)\left\{\left(1-n\right)c_{sc}^{2}+2\alpha\left(n+1\right)\right\}}{\left\{\left(1-n\right)c_{sc}^{2}+\alpha\left(n+1\right)\right\}^{2}}\right],$$
\begin{equation}\label{accretioncqg12}
\mathcal{B}=-\frac{2}{c_{sc}^{4}}\frac{\left(c_{sc}^{2}-\alpha\right)\left(n+1\right)\left\{\left(1-n\right)c_{sc}^{2}+2\alpha\left(n+1\right)\right\}}{\left\{\left(1-n\right)c_{sc}^{2}+\alpha\left(n+1\right)\right\}}\left[F_{g}(x_{c})-\frac{\lambda^{2}}{x_{c}^{3}}\right],
\end{equation}

\begin{equation}
\mathcal{C}=\left\{\frac{3\lambda^{2}}{x_{c}^{4}}-\left(\frac{dF_{g}}{dx}\right)_{x=x_{c}}\right\}-\left[\left\{\frac{1}{F_{g}}\left(\frac{dF_{g}}{dx}\right)^{2}\right\}_{x=x_{c}}-\frac{3}{x_{c}^{2}}-\left(\frac{1}{F_{g}}\frac{d^{2}F_{g}}{dx^{2}}\right)_{x=x_{c}}\right]\frac{u_{c}^{2}}{2}~~~~~~~~~~~~~~~~~~~~~~~~~~~~~~
$$$$~~~~~~~~~~~~~~~~~~~~~~~~~~~~~~~~~~~~~~~~-\frac{u_{c}^{2}}{2c_{sc}^{8}}\left[\left(c_{sc}^{2}-\alpha\right)\left(n+1\right)\left\{\left(1-n\right)c_{sc}^{2}+2\alpha\left(n+1\right)\right\}\right]\left[F_{g}(x_{c})-\frac{\lambda^{2}}{x_{c}^{3}}\right]^{2},
\end{equation}
where $x_c$ is the critical point where both the denominator and the numerator of the equation~(\ref{eqn:diffeqn:fluid}) turn zero and $u_c$ is the value of radial velocity at $x=x_c$ and $c_{sc}$ is the speed of sound at $x=x_c$.

In the next section, we will find out the radial velocity solutions for different parametric values and physically interpret them accordingly.
\section{Results and Analysis}
\label{sec:result:analysis}

Let us take the adiabatic fluid $p=K\rho^\Gamma$  to accrete on the concerned BH at first.

We find out the values of radial velocities for different radial distances for adiabatic ($\Gamma = 1.6$) accretion and for $a=0$, and plot them in figures 1.1.(a) to 1.1.(e). In every plot, the log of the radial distances (in $GM/c^2$ unit) from the BH is taken along the  horizontal axis and the radial speed of the accreting fluid (in $c$ unit) is taken along the vertical axis. Fig. 1.1.(a)- 1.1.(e) are for non-rotating BHs. Similarly, 1.2.(a)-1.2.(e), 1.3.(a)-1.3.(e), 1.4.(a)-1.4.(e) and 1.5.(a) - 1.5.(e) are for rotational parameters a=0.1, 0.5 , 0.9 and 0.998 respectively (i.e., four sets of curves are for rotating BHs.) In every set, the graphs marked by (a) is drawn for $\mathcal{A}_q=0, \omega_q=0$ i.e., no quintessential effect is considered inside the BH.  Graphs with tag (b) are for $\omega_q= 1/3$, i.e., we have taken our BH to reside inside radiation enriched background. The graphs, which are with label (c) do consider $\omega_q=0$, i.e., the concerned BH is residing in the pressure less dust background. The case where BH is dipped in quintessence with $\omega_q = -2/3$, is considered in the plots with tag (d). Finally, at phantom boundary $\omega_q = -1$, the nature of the graphs are incorporated in graphs with label (e).

In every figure, we see two curves: one solid and the other dotted. The solid curve depicts the inward falling speed , i.e., the accretion speed. The dotted curve, on the other hand, signifies the wind speed, i.e., the speed with which the fluid is thrown away from the disc. Accretion curves show very small radial inward speed if we are very far from BH. As we approach towards the BH, the speed increases and near to the BH, it shows a speed equal to that of light. Wind, when near to the BH, shows very small speed. Then we move far from the BH and the speed increases. Up to a distance, the rate of increment of speed  is high and then the rate falls and the wind speed beyond that distance becomes almost constant or slowly decreasing.

Fig 1.1.(a) depicts the accretion-wind curves for non-rotating, non-quintessence case. The accretion curve is increasing towards the BH and the wind is small near to the   horizon of BH and increases as we move far from the BH and becomes almost constant.

While plotting figure 1.1.(b), we have taken $x_c$ to occur at 13 unit distance. Then at the critical point accretion-wind speed is found to be equal critical fluid speed = 0.228622 and critical sound speed=0.260669. Wind speed does not rise much after  we cross the critical point. This signifies that the accretion dominates highly if $\omega_q = 1/3$. Radiation induced BH attracts much. Even at a distance of 13 unit, the critical point  can be formed, i.e., even at a higher distance, the attractive force is strong enough to balance the outward wind pressure. Another important thing is to be noted that the accretion is possible even at or after the distance of $10^3$ units. So, the attractive nature is even working at such a high distance far from the central engine.

Figure 1.1.(c) is for adiabatic accretion onto a BH which is embedded in a pressureless dust background (i.e. $\omega_q = 0$). Firstly, for this case, to obtain physical curves, we had to shift the critical point towards the BH (at a distance of 10 unit) and the resulting accretion-wind speed at $x_c, u_c$ is higher that the case of $\omega_q= 1/3$. The wind speed, before being constant valued, gets increased up to a certain range $x_c+h_1$. In this case, the wind is stronger than the $\omega_q=1/3$ case and the dominance of accretion is lesser. Non-zero accretion speed is possible to find at a distance of thousands units though the accretion speed falls little more rapidly than the $\omega_q=1/3$ case. So, we see the starting of the trends that (i) to balance the wind we have to come towards the BH, (ii) $u_c$ is higher, (iii) accretion speed becomes weaker at the scale of $10^2$ than the $\omega_q = 1/3$ case and  (iv) wind becomes stronger. Even when the DE effect inside the BH is not incorporated, only the radiation background is changed to a pressure less dust, we get stronger wind and weaker accretion. This, on the other hand, speculates  that the BH's attracting force is reduced when $\omega_q$ is reduced.

Figure 1.1.(d) considers BH to be embedded in the quintessence with $\omega_q=-2/3$. The critical point is shifted to $x_c=8$ units. $u_c$ raised higher than the $w_q=0$ case. Wind increases up to $x_c + h_2$ ($h_2 >h_1$) and then becomes almost constant (slowly increasing actually). Accretion reduces down to very small values around $50$ unit. Though a small accretion could be found even at thousands units of distance. It can be predicted that if a BH is embedded in a quintessence universe its attracting power becomes lesser even if the matter accreting on it is adiabatic type, i.e., following the EoS $p=K \rho^\Gamma$. As a result, accretion becomes weaker and wind starts to become stronger than the $\omega_q=1/3$ or $\omega_q=0$ cases. The feeding process for a BH turns to be weaker in quintessence universe. As quintessence or DE is a global phenomenon, not particularly a local one, it might have a lesser effect on a local incidence like accretion. This motivates us to take a smaller value ($~10^{-10}$) for the regulating parameter $\mathcal{A}_q$. While  we have chosen $\omega_q=-2/3$ and $\omega_q=-1$, with even such a small effect, we observe, that the wind curve to rise higher and accretion to depress at a nearer region to the BH. This indicates the BH's attraction substantially reduced by the quintessence interface.

\begin{figure}[H]
	\centering
	\renewcommand\thesubfigure{(\alph{subfigure})}
		\subfigure[$x=13,\omega_q=0$]{\includegraphics[width=0.19\textwidth]{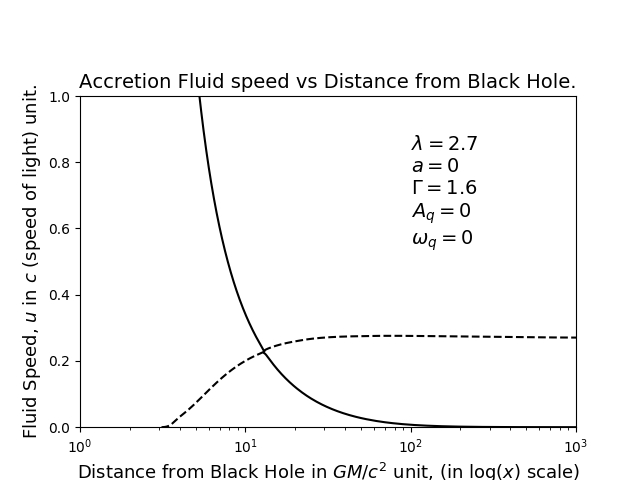}}
		\subfigure[$x=13,\omega_q=1/3$]{\includegraphics[width=0.19\textwidth]{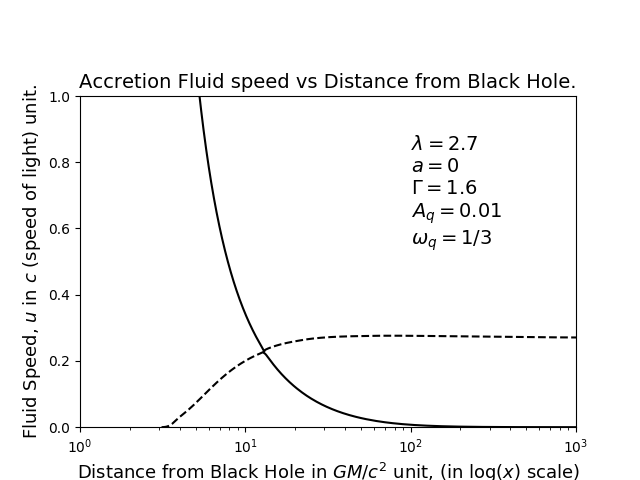}} 
		\subfigure[$x=10,\omega_q=0$]{\includegraphics[width=0.19\textwidth]{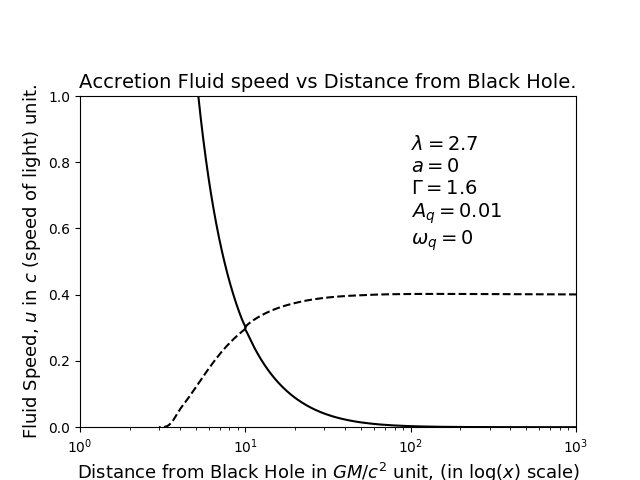}}
	\subfigure[$x=8,\omega_q=-2/3$]{\includegraphics[width=0.19\textwidth]{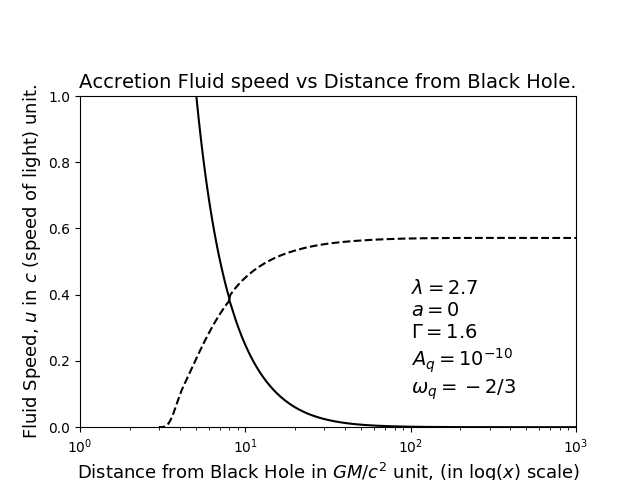}}
	\subfigure[$x=6,\omega_q=-1$]{\includegraphics[width=0.19\textwidth]{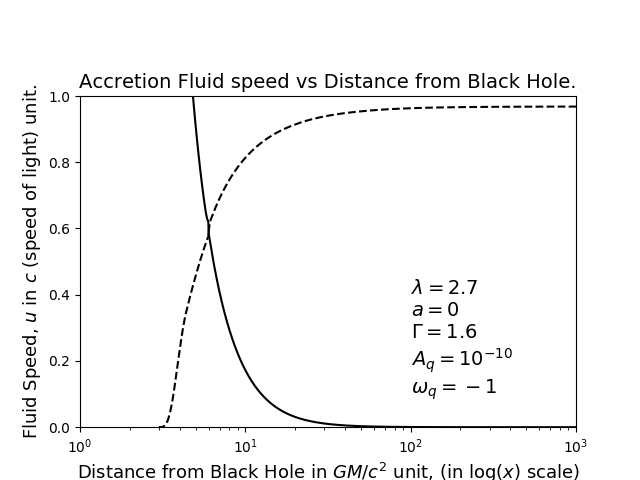}}
	\caption*{\textbf{\emph{Figure 1.1:}} Images for $a=0.0$}

    \setcounter{subfigure}{0}

	\subfigure[$\omega_q=0$]{\includegraphics[width=0.19\textwidth]{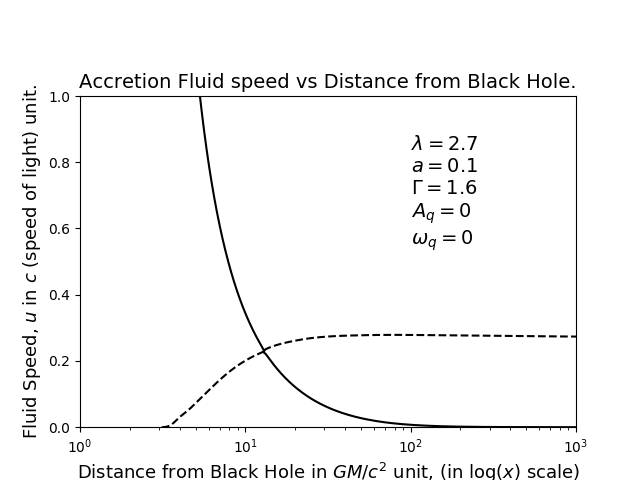}}
	\subfigure[$\omega_q=1/3$]{\includegraphics[width=0.19\textwidth]{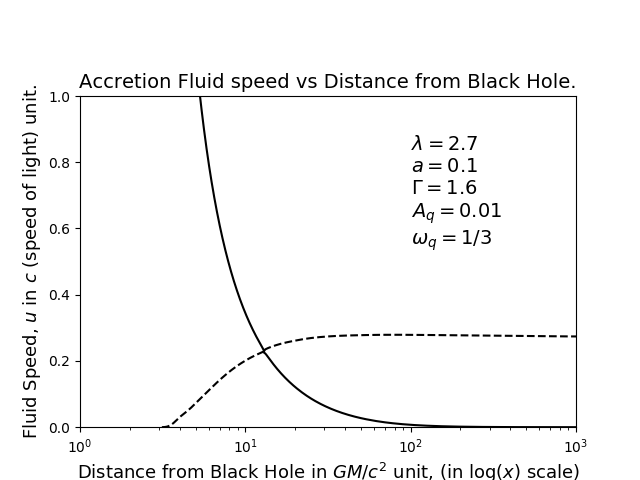}} 
	\subfigure[$\omega_q=0$]{\includegraphics[width=0.19\textwidth]{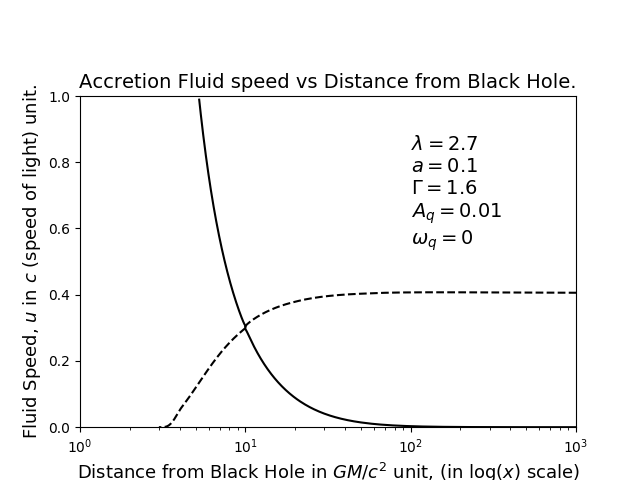}} 
	\subfigure[$\omega_q=-2/3$]{\includegraphics[width=0.19\textwidth]{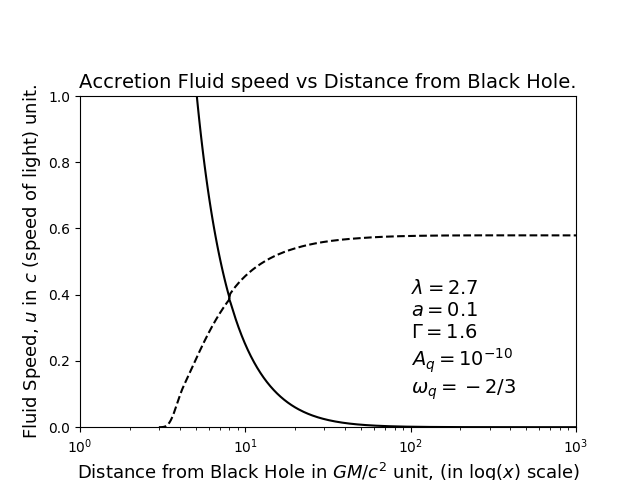}}
	\subfigure[$\omega_q=-1$]{\includegraphics[width=0.19\textwidth]{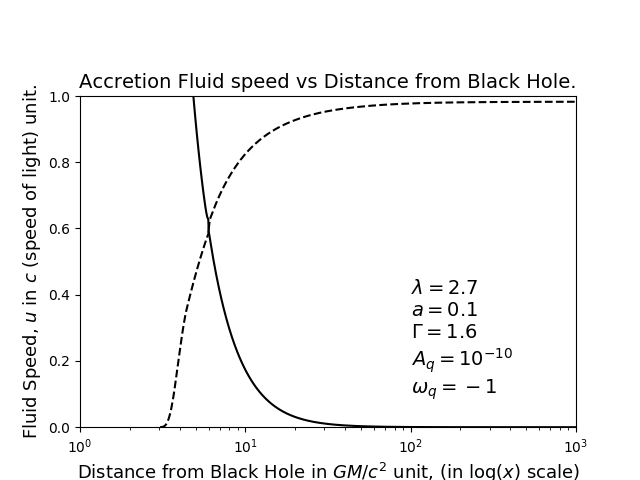}}
	\caption*{\textbf{\emph{Figure 1.2:}} Images for $a=0.1$}
	\setcounter{subfigure}{0}
	
	\subfigure[$\omega_q=0$]{\includegraphics[width=0.19\textwidth]{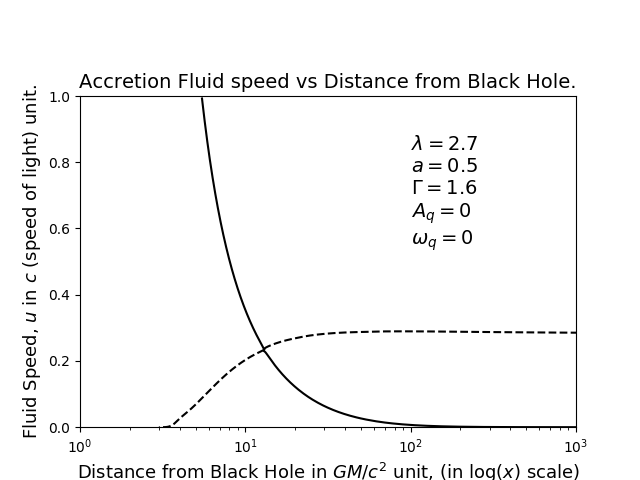}} 
	\subfigure[$\omega_q=1/3$]{\includegraphics[width=0.19\textwidth]{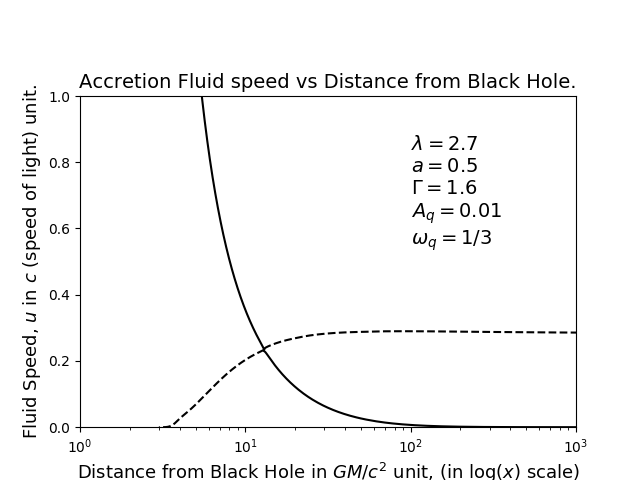}} 
	\subfigure[$\omega_q=0$]{\includegraphics[width=0.19\textwidth]{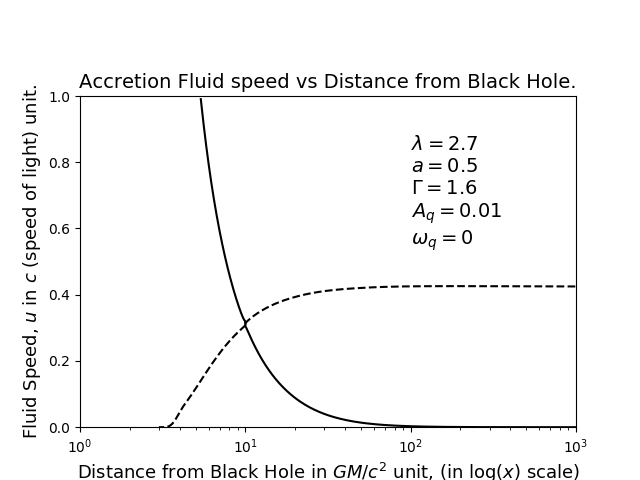}} 
	\subfigure[$\omega_q=-2/3$]{\includegraphics[width=0.19\textwidth]{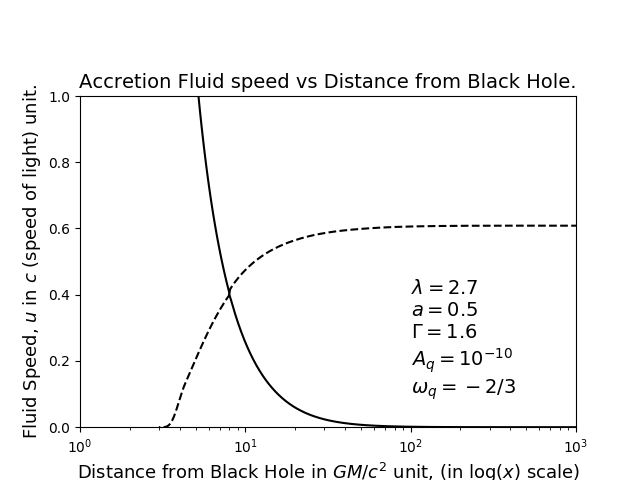}}
	\subfigure[$\omega_q=-1$]{\includegraphics[width=0.19\textwidth]{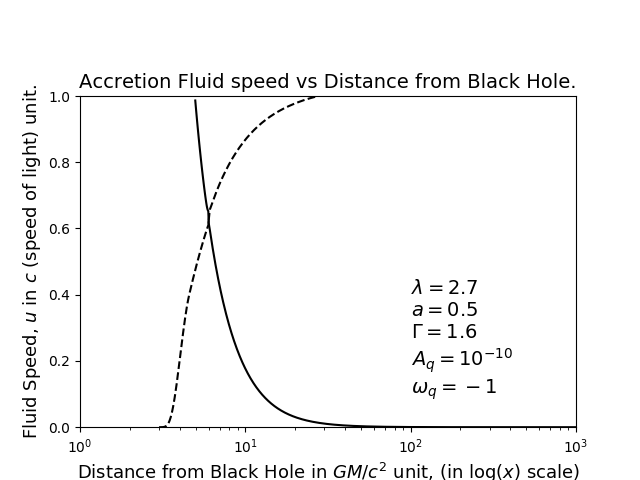}}
	\caption*{\textbf{\emph{Figure 1.3:}} Images for $a=0.5$}
	\setcounter{subfigure}{0}
	\subfigure[$\omega_q=0$]{\includegraphics[width=0.19\textwidth]{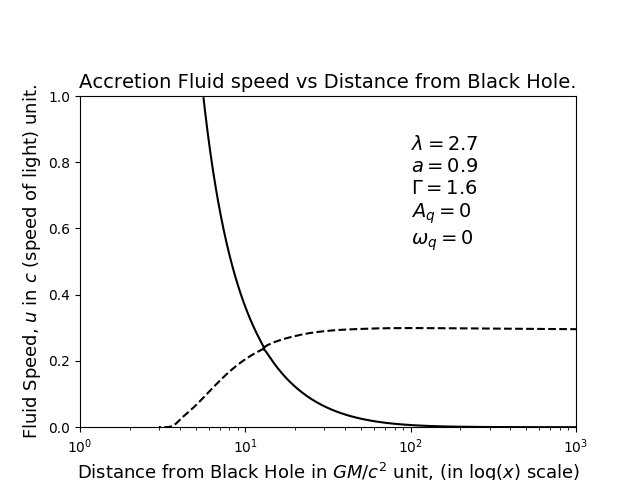}}
	\subfigure[$\omega_q=1/3$]{\includegraphics[width=0.19\textwidth]{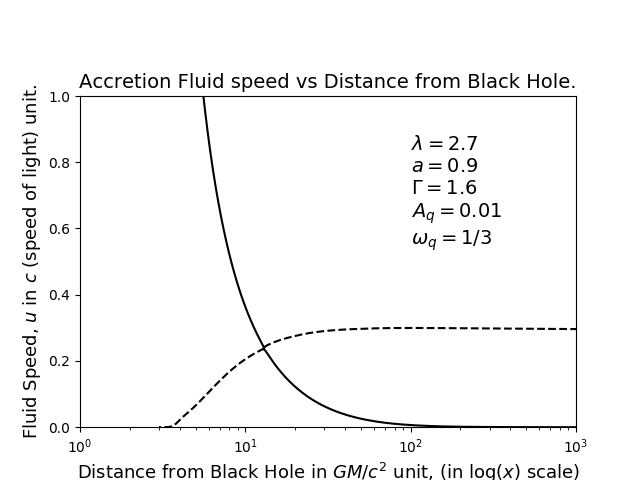}} 
	\subfigure[$\omega_q=0$]{\includegraphics[width=0.19\textwidth]{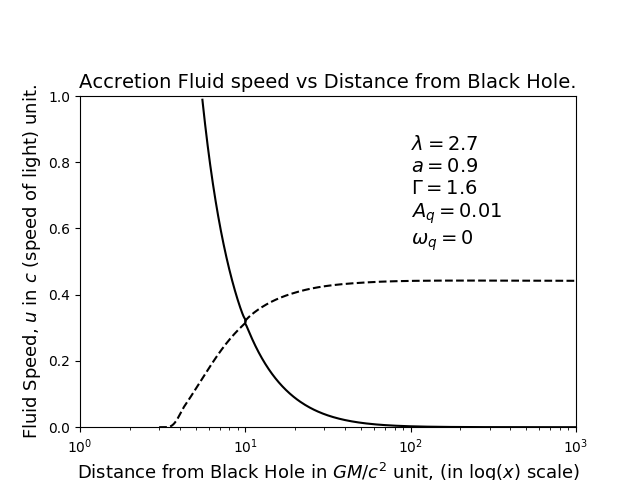}} 
	\subfigure[$\omega_q=-2/3$]{\includegraphics[width=0.19\textwidth]{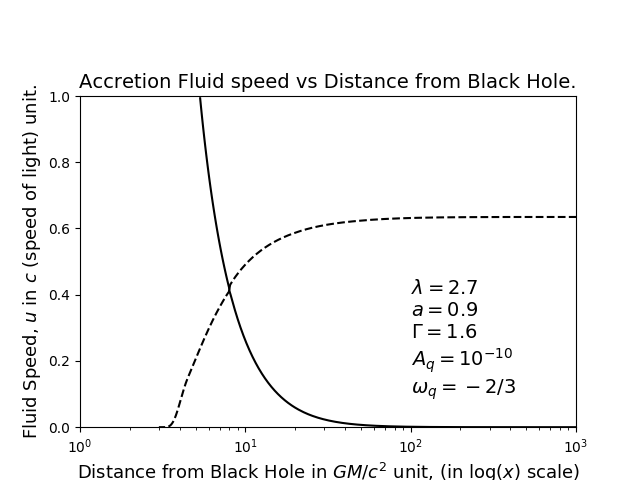}}
	\subfigure[$\omega_q=-1$]{\includegraphics[width=0.19\textwidth]{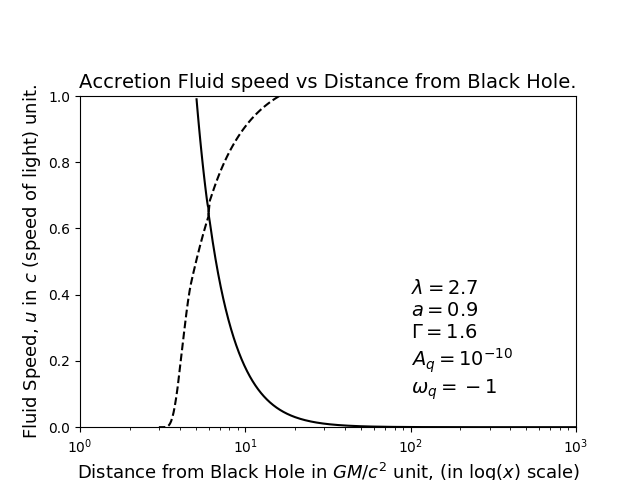}}
	\caption*{\textbf{\emph{Figure 1.4:}} Images for $a=0.9$}
	\setcounter{subfigure}{0}
	\subfigure[$\omega_q=0$]{\includegraphics[width=0.19\textwidth]{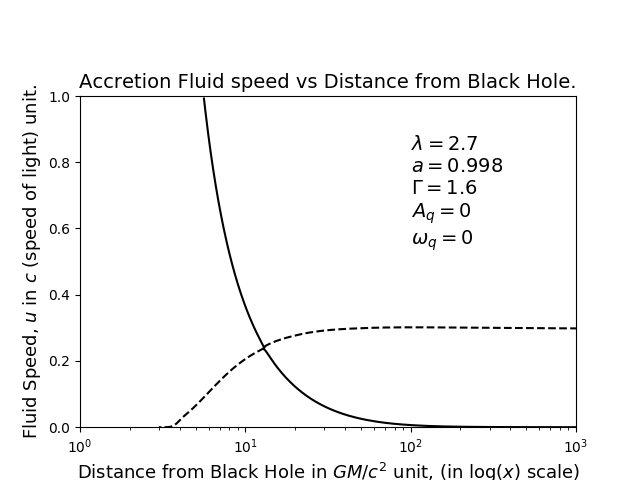}}
	\subfigure[$\omega_q=1/3$]{\includegraphics[width=0.19\textwidth]{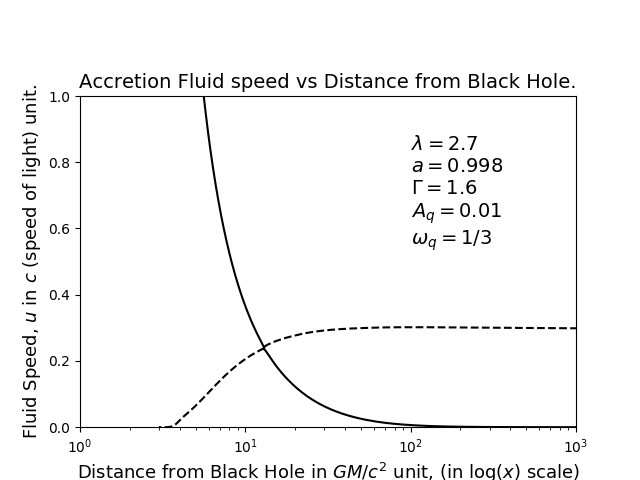}} 
	\subfigure[$\omega_q=0$]{\includegraphics[width=0.19\textwidth]{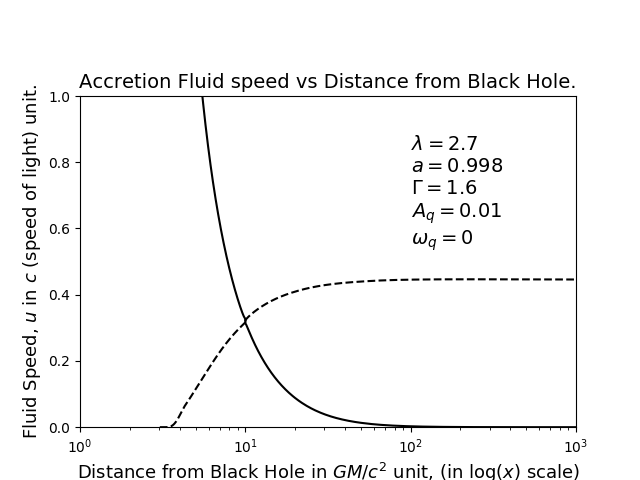}} 
	\subfigure[$\omega_q=-2/3$]{\includegraphics[width=0.19\textwidth]{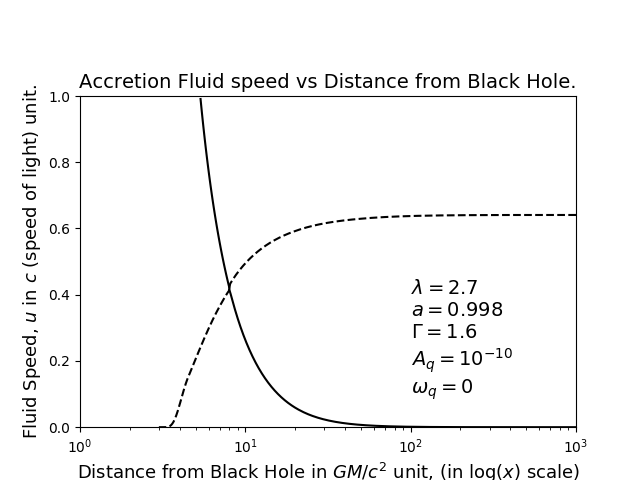}}
	\subfigure[$\omega_q=-1$]{\includegraphics[width=0.19\textwidth]{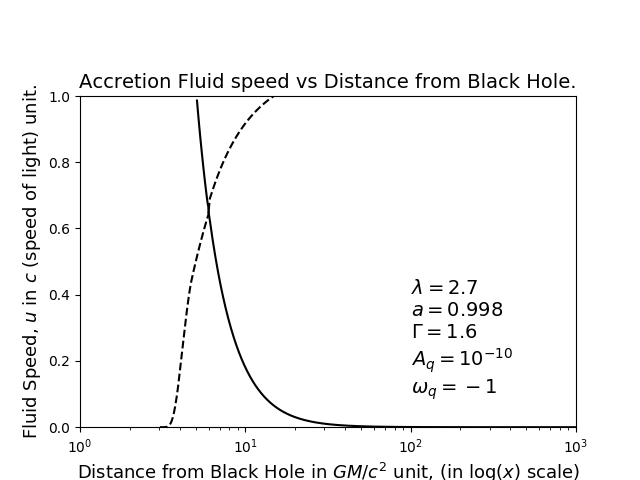}}
	\caption*{\textbf{\emph{Figure 1.5:}}Images for $a=0.998$}
	\caption{Accretion and Wind fluid radial  speeds for different spin and quintessence parameters. Adiabatic fluid is taken to be the accreting agent.}
	\label{fig:x:u}
\end{figure}

\begin{figure}[H]
	\centering
	\setcounter{subfigure}{0}
	\subfigure[$\omega_q=0$]{\includegraphics[width=0.19\textwidth]{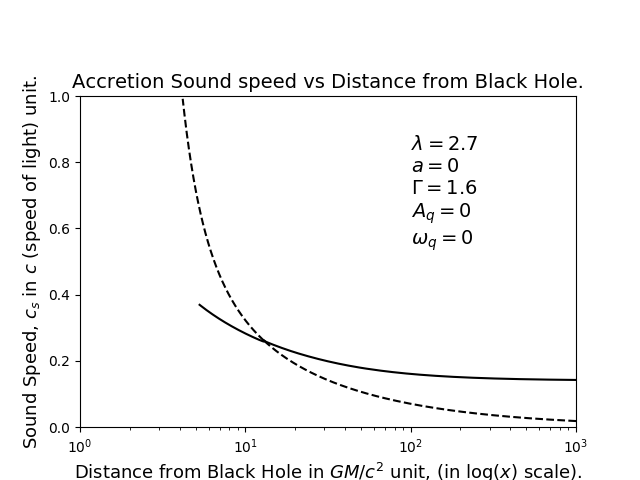}}
	\subfigure[$\omega_q=1/3$]{\includegraphics[width=0.19\textwidth]{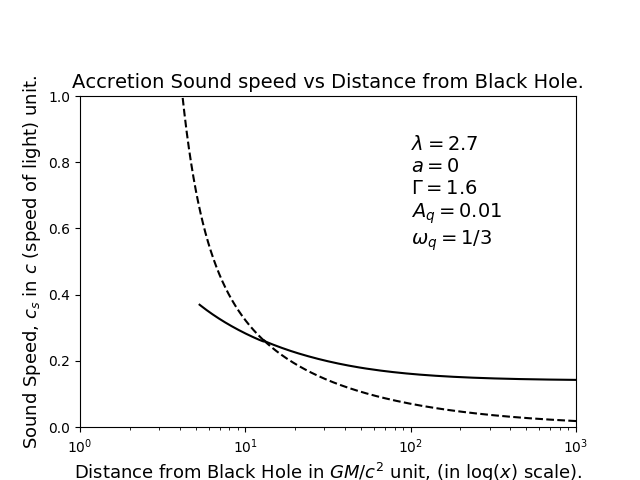}} 
	\subfigure[$\omega_q=0$]{\includegraphics[width=0.19\textwidth]{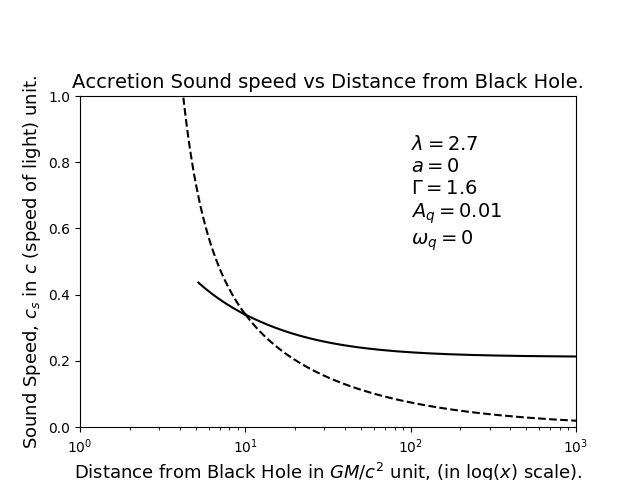}} 
	\subfigure[$\omega_q=-2/3$]{\includegraphics[width=0.19\textwidth]{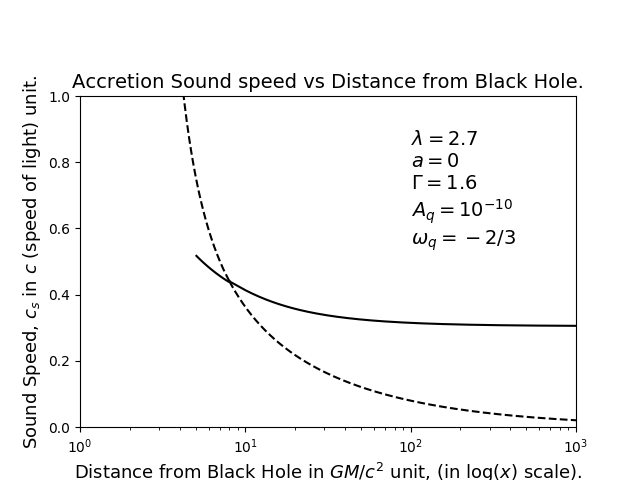}}
	\subfigure[$\omega_q=-1$]{\includegraphics[width=0.19\textwidth]{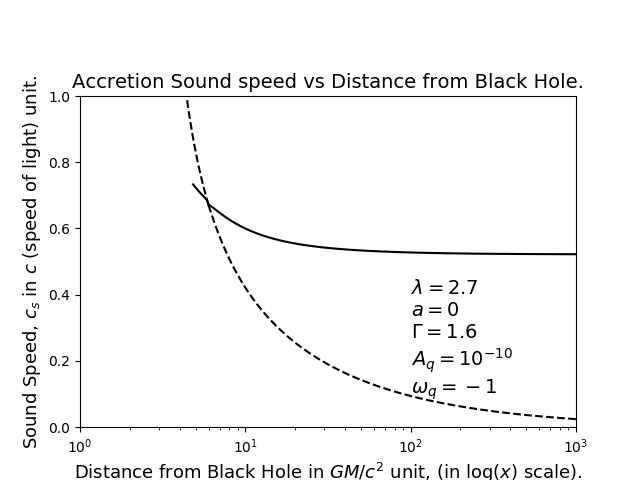}}
	\caption*{\textbf{\emph{Figure 2.1:}} Images for $a=0.0$}
	\setcounter{subfigure}{0}
	\subfigure[$\omega_q=0$]{\includegraphics[width=0.19\textwidth]{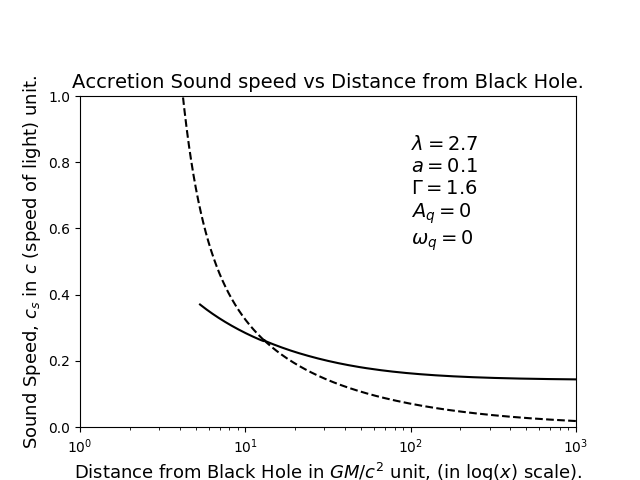}}
	\subfigure[$\omega_q=1/3$]{\includegraphics[width=0.19\textwidth]{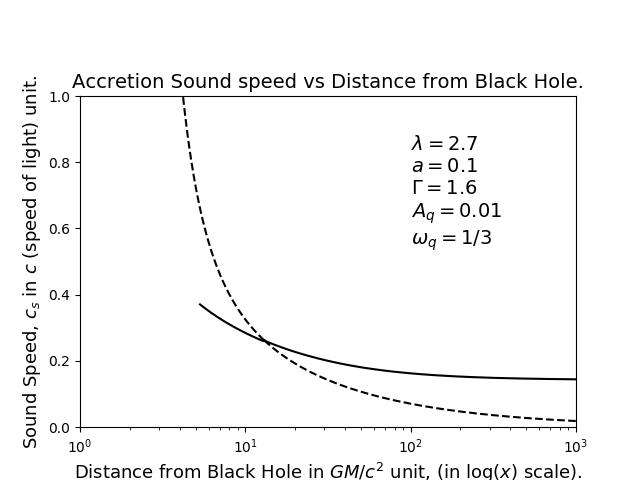}} 
	\subfigure[$\omega_q=0$]{\includegraphics[width=0.19\textwidth]{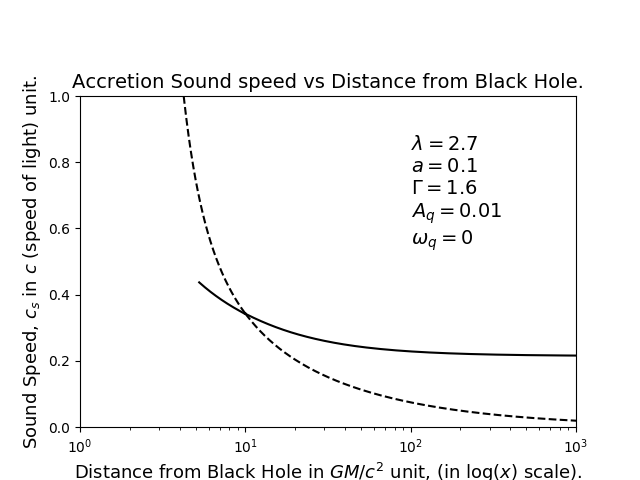}} 
	\subfigure[$\omega_q=-2/3$]{\includegraphics[width=0.19\textwidth]{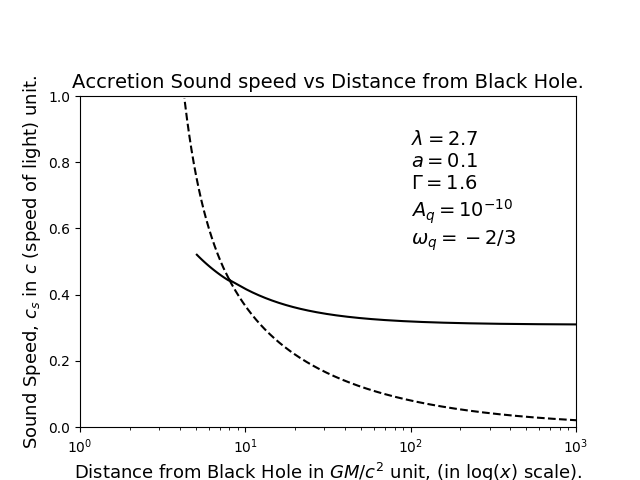}}
	\subfigure[$\omega_q=-1$]{\includegraphics[width=0.19\textwidth]{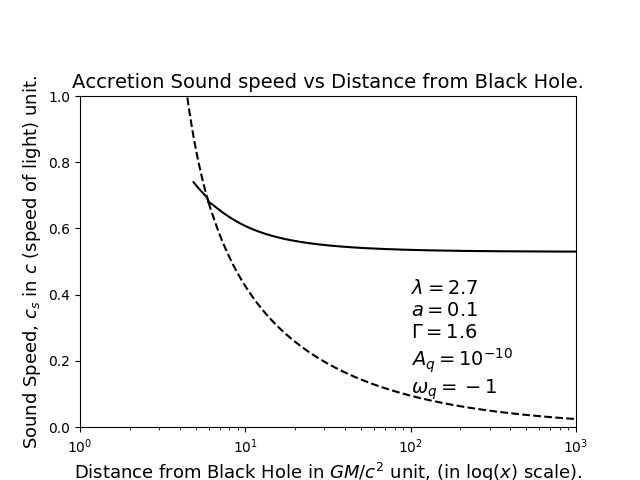}}
	\caption*{\textbf{\emph{Figure 2.2:}} Images for $a=0.1$}
	\setcounter{subfigure}{0}
	\subfigure[$\omega_q=0$]{\includegraphics[width=0.19\textwidth]{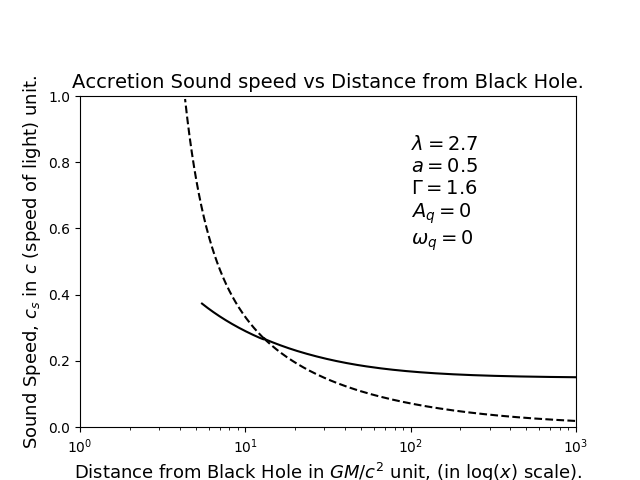}} 
	\subfigure[$\omega_q=1/3$]{\includegraphics[width=0.19\textwidth]{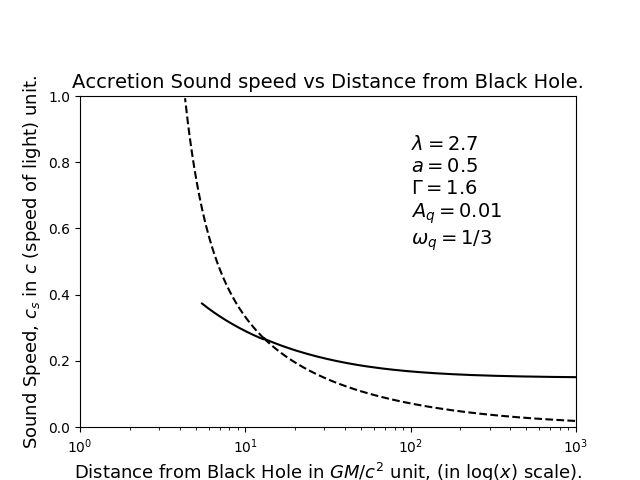}} 
	\subfigure[$\omega_q=0$]{\includegraphics[width=0.19\textwidth]{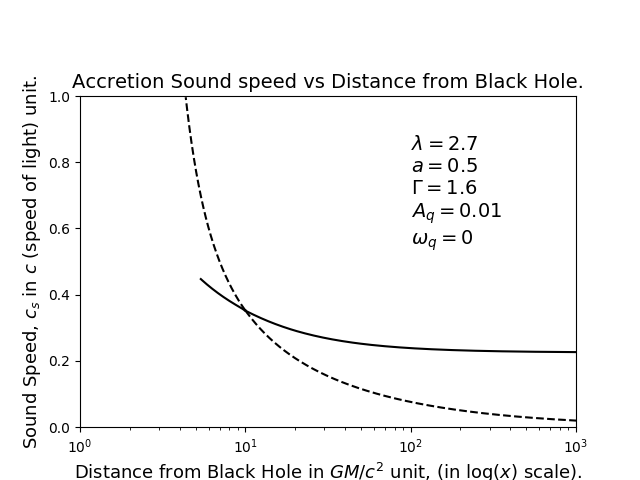}} 
	\subfigure[$\omega_q=-2/3$]{\includegraphics[width=0.19\textwidth]{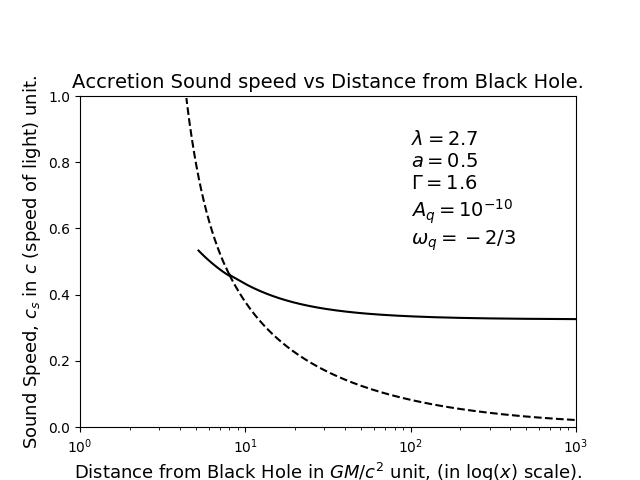}}
	\subfigure[$\omega_q=-1$]{\includegraphics[width=0.19\textwidth]{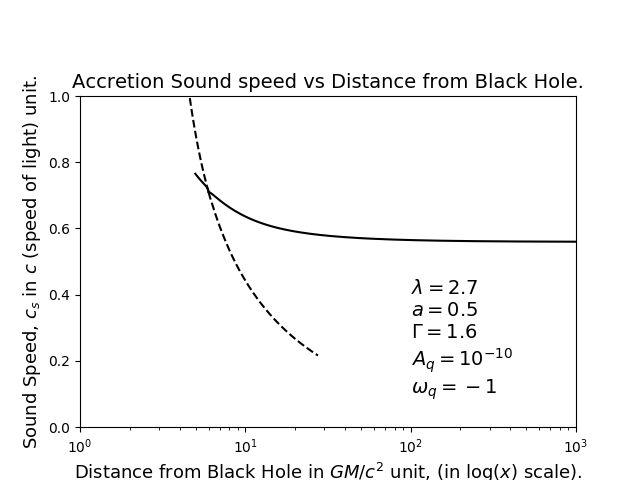}}
	\caption*{\textbf{\emph{Figure 2.3:}} Images for $a=0.5$}
	\setcounter{subfigure}{0}
	\subfigure[$\omega_q=0$]{\includegraphics[width=0.19\textwidth]{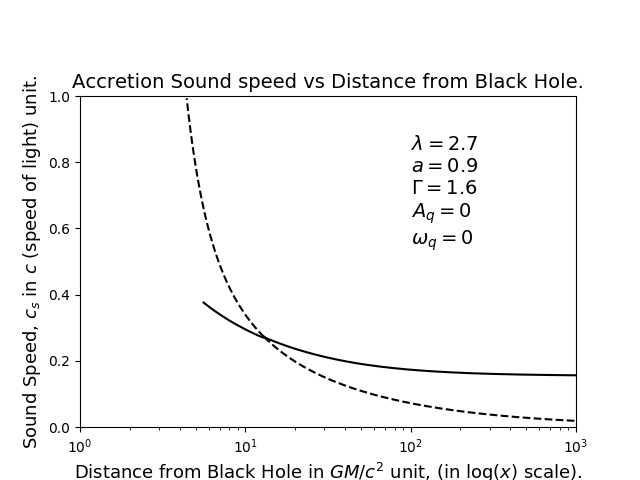}}
	\subfigure[$\omega_q=1/3$]{\includegraphics[width=0.19\textwidth]{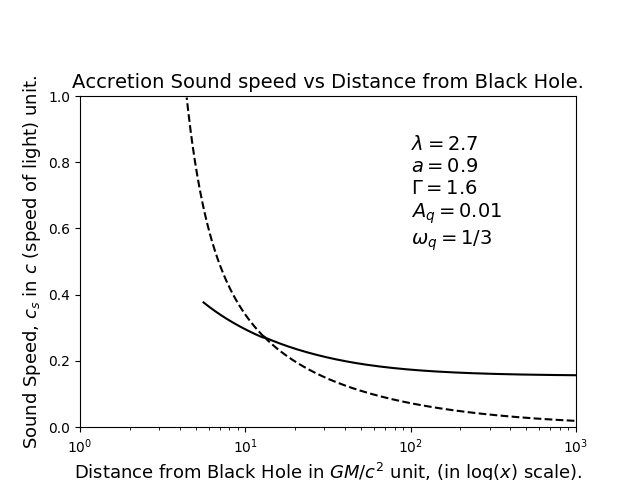}} 
	\subfigure[$\omega_q=0$]{\includegraphics[width=0.19\textwidth]{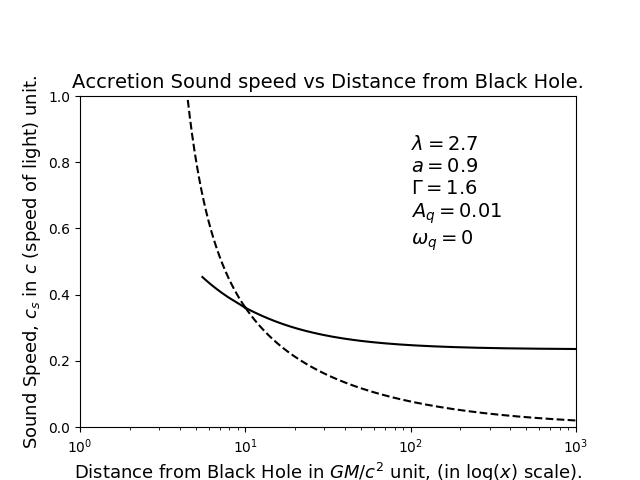}} 
	\subfigure[$\omega_q=-2/3$]{\includegraphics[width=0.19\textwidth]{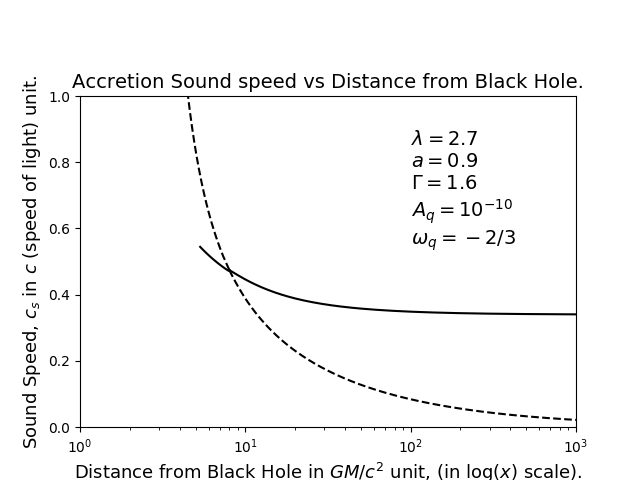}}
	\subfigure[$\omega_q=-1$]{\includegraphics[width=0.19\textwidth]{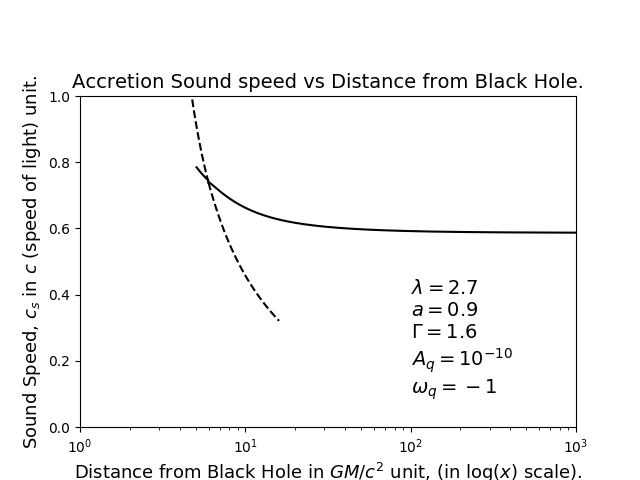}}
	\caption*{\textbf{\emph{Figure 2.4:}} Images for $a=0.9$}
	\setcounter{subfigure}{0}
	\subfigure[$\omega_q=0$]{\includegraphics[width=0.19\textwidth]{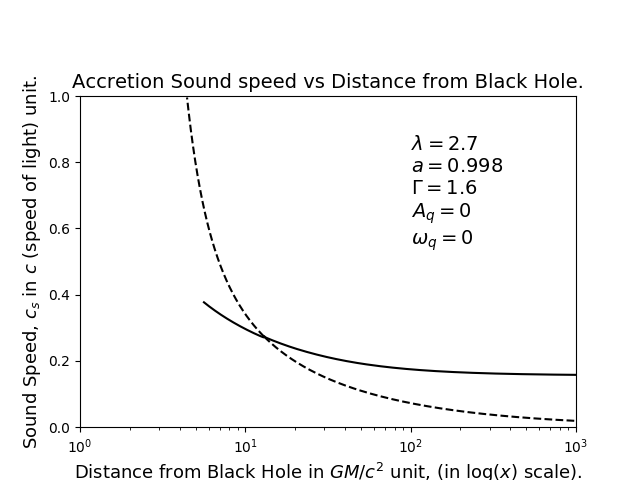}}
	\subfigure[$\omega_q=1/3$]{\includegraphics[width=0.19\textwidth]{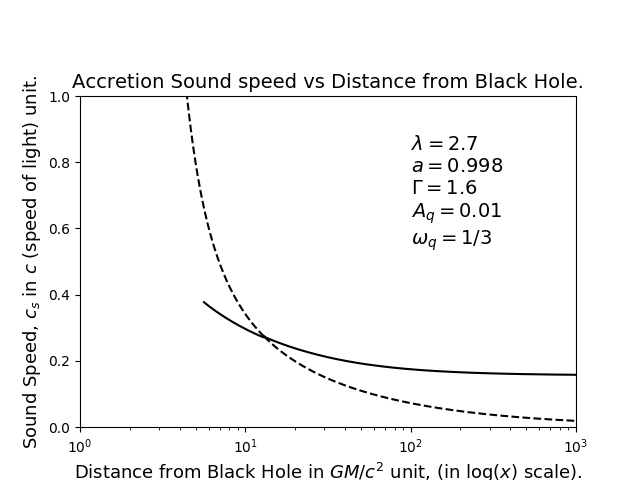}} 
	\subfigure[$\omega_q=0$]{\includegraphics[width=0.19\textwidth]{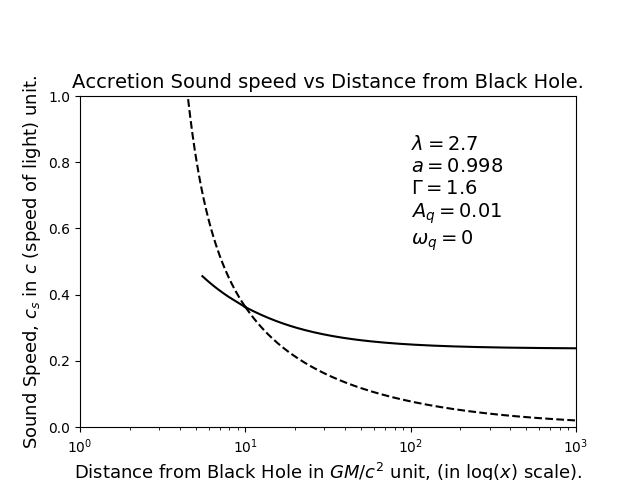}} 
	\subfigure[$\omega_q=-2/3$]{\includegraphics[width=0.19\textwidth]{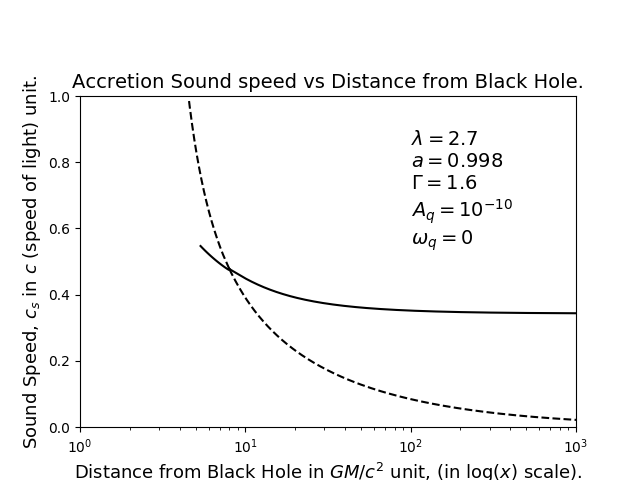}}
	\subfigure[$\omega_q=-1$]{\includegraphics[width=0.19\textwidth]{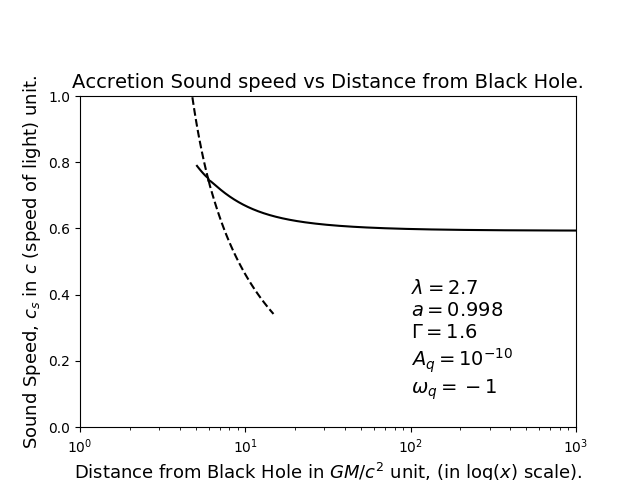}}
	\caption*{\textbf{\emph{Figure 2.5:}} Images for $a=0.998$}
	\caption{Accretion and Wind sound speeds for different parameters for adiabatic gas as accreting fluid.}
	\label{fig:x:cs}
\end{figure}

\begin{figure}[h]
	\centering
	\setcounter{subfigure}{0}
	\subfigure[$\omega_q=-1$]{\includegraphics[width=0.32\textwidth]{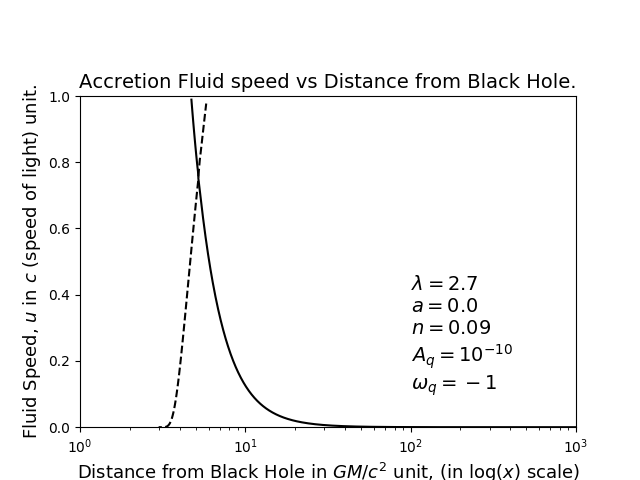}}
	\subfigure[$\omega_q=-1$]{\includegraphics[width=0.32\textwidth]{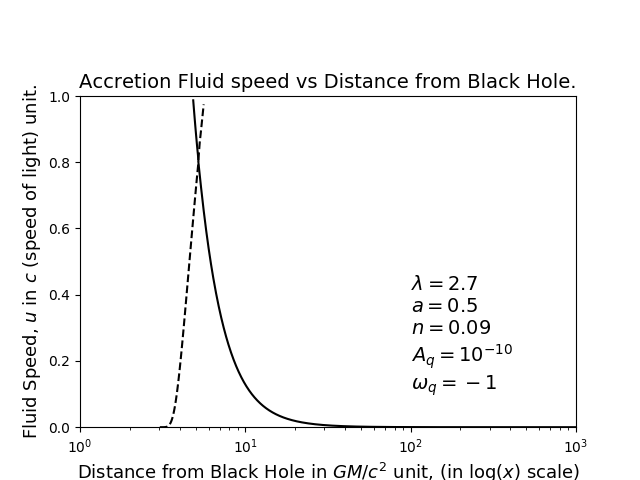}} 
	\subfigure[$\omega_q=-1$]{\includegraphics[width=0.32\textwidth]{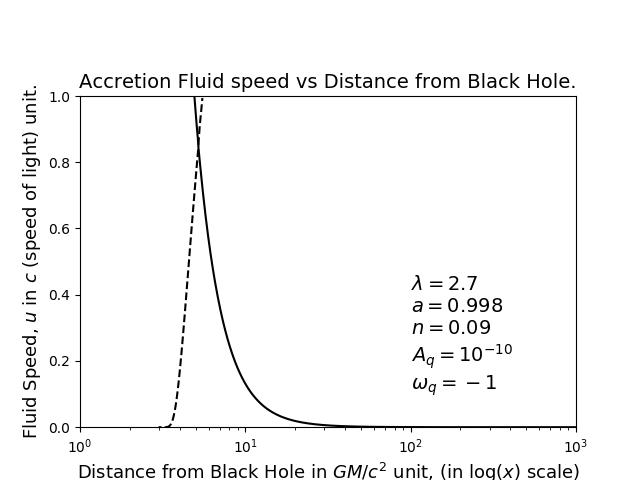}} 
	\caption{Accretion and Wind fluid radial speeds for different parameters for modified Chaplygin gas as accreting fluid.}
	\label{img:de:fluids}
\end{figure}

\pagebreak

As we shift towards the $\omega_q = -1$ case, i.e., we stand on the phantom barrier in the figure 1.1.(e), we need to shift the critical point to inner region, i.e., up to 6 unit distance and as a result, the critical point values of accretion wind speed raise higher than the $\omega_q = -2/3$ case. Wind speed grows high up to a radial distance of $x_c+h_3$ ($h_3 >h_2 > h_1$) and then becomes slowly increasing. Value of the accretion speed falls down at a nearer radial distance than the $\omega_q = -2/3$ case. However, the wind speed increments tradition is kept the change is much as we go form $\omega_q=-2/3$ to $\omega_q = -1$ case. So, at a near phantom regime, the BH's attraction as an attracting engine falls abruptly than before. During this analysis, the BH is not rotating and the nature of accreting fluid is taken to be adiabatic type. So DE, only staying inside the BH, is weakening the attracting power of the central engine, depressed the speed of accretion and catalyte the speed of Wind.

We will now concentrate on the figures~\ref{fig:x:u}.2.(a)~-~\ref{fig:x:u}.2.(e) which are the $a=0.1$ cases for the \ref{fig:x:u}.1.(a)~-~\ref{fig:x:u}.1.(e). The natures are respectively the same. The only clear difference is observed for $\omega_q = -1$ case. The increased rotation of the BH increases the wind speed. This supports the fact that the rotational speed inhibits the decreasing power of attraction of a DE induced BH. As a result, the wind becomes stronger than the non-rotating case.

Figure~\ref{fig:x:u}.3.(a)~-~\ref{fig:x:u}.3.(e) are cases for $a=0.5$. Here, for all the figures, we observe the wind become stronger than the corresponding $a=0.1$ case. So, the trend of weakening of the accretion and strengthening of the wind is carried on. Most interesting thing happens for the $\omega_q = -1$ case. At a finite distance, ( $x=x_{\omega 0.5}$ say ), the wind speed becomes equal to that of light. So, if $x > x_{\omega 0.5}$, the speed is so high, that the matters are thrown out with a speed which is higher than the in-falling speed. This confirms the weakening of the accretion disc and hence the feeding process of the BH.

Figures~\ref{fig:x:u}.4.(a)~-~\ref{fig:x:u}.4.(e) are for $a=0.9$ and  Figures~\ref{fig:x:u}.5.(a)~-~\ref{fig:x:u}.5.(e) are for $a=0.998$ cases. The common nature is followed. $\omega_q = -1$ cases show the wind speed to be equal to that of light at a finite distance (say $x=x_{\omega 0.9}$ for $a=0.9$ and $x=x_{\omega 0.998}$ for $a=0.998$), and that  $x_{\omega 0.998} < x_{\omega 0.9} < x_{\omega 0.5}$. Along with this, $\frac{du}{dx}|_{x_{\omega 0.998}} >\frac{du}{dx}|_{x_{\omega 0.9}} >\frac{du}{dx}|_{x_{\omega 0.5}}$, i.e., for high BH rotation, the wind speed takes the speed of light with higher slope.

We plot the sound speed through accreting and wind fluid vs the log of distance from the BH in the figure~\ref{fig:x:cs}.1.(a) through~\ref{fig:x:cs}.5.(e). The accretion sound speed is given by solid lines and for wind, we have used dashed lines. Accretion sound speed stays in a finite region of values whereas the wind sound speed is high when we are very near to the BH and it falls quickly as we move far from the BH. Inclusion of DE in the BH solution does not hamper the slope pattern of the wind sound speed(figure~\ref{img:de:fluids}(a)-figure~\ref{img:de:fluids}(c)). Only outer portion is truncated if we involve more negative energy and increase the rotation of the BH. Value of accretion sound speed increases as we increase the rotation or the negativity of the DE involved. It is clear, that the sound speed increases very slowly in the accretion flow whereas the fluid accretion velocity increases rapidly. For wind sound, speed increases towards  BH but fluid sound velocity decreases. The rate of outward fall in sound speed is higher for high rotation and DE effect. For fluid velocity, this is done for the outward hike in values.

\section{Observational Supports for the Theoretical Results}
\label{sec:observational:support}

Major merger of gas-rich galaxies do full up quasars at high red shift and hence most massive SMBHs are grown. At lower red shift, other portions of fueling mechanisms and fainter luminosities are likely to take place~\cite{marulli:2007}. Distribution of a  BH's mass can be represented as redshift's function where a continuity equation, which describes the `flow' of BH number density is employed~\cite{small:blandford:1992}. Quasar luminosity function traces the accretion of matter onto BHs modulo bolometric correction and accretion effectively  and so it is now as a constraint on the rate of change in the mass density of SMBH. Authors of reference~\cite{yu:lu2008} points out that the early type galaxies dominate local BH  mass function  at $M_{BH} \geq 4\times 10^7M_\odot$. According to the reference~\cite{Kelly:2010}, for $z\ge 2$, the SDSS quasar sample is only around $10\%$ complete at $M_{BH}~10^9 M_\odot$ and becomes incomplete at lower masses. In the reference~\cite{Hopkins:2006}, it was shown that the rate of mass increment is related to the time as $\dot{M}(t) \propto t^{-\beta}$ which indicates as we go towards the future $\dot{M}(t)$ reduces. It is established that the evolution of cosmological SMBH mass density for broad line quasars tracks evolution in the cosmological accretion rate density of SMBHs~\cite{soltan:1982}\cite{marconi:2004}.  Summarizing  these results for Sloan Digital Sky  Survey (SDSS) Data Release 3, using nearly 15,180 quasars over $0.3<z<5$, it is clear that the matter accretion rate decreased over time. The best feeding process for a SMBH seed was not done at least in present time. The reference~\cite{10:1093:mnras:sts261} also shows that the peak formation rate does decrease with redshift from 23 percent at $z=0$ to 9 percent at $z=4$ along with an increase of halo mass from $10^{11.8}M_\odot$ to $10^{12.5}M_\odot$. Thinking time wise, we can follow the decrease in mass.

Super critical radiatively inefficient accretion on high redshift quasars are happened to be occurred at significantly early epochs i.e. $z\geq 10$. Where as we find radiatively efficient accretions on quasars [like ULASJ11200(z=7.09)], SDSS J0100 (z=6.3), SDSSJ1148 (z=6.41)]~\cite{Trakhtenbrot:2017}. These incidents supports the fact that accretion in earlier epochs where stronger than the near redshift phases.

Beside the accretion weakening, we must try to find out some evidences for wind strengthening. The measurement of the event horizon of a $10^9M_\odot$ SMBH is approximately equal to $10^{15}$ cm, which is a billionth of a typical galactic bulge. Even the sphere of BH's gravitational influence is thousand times smaller than the galactic bulge's size. This is supposed to get formed of a disc outflow. A powerful AGN accretion disc having a relativistic velocity of $0.25c$ in x-ray spectrum of IRASF11119+3257 is found. This is a nearly ($z=0.189$) optically classified type 1 and a powerful molecular outflow is also observed~\cite{Veilleux:2013}.

On the other hand, we can find another quasar APM08279+5255 with 0.02c~\cite{Chartas:2009} outflow velocity and which is situated at $z=3.911$. Like this, as we look nearer, we find a higher wind velocity. We extrapolate these results to conclude that with increment in redshift, wind speed falls.

\section{Brief Discussions and Conclusions}
\label{sec:conclusion}

In this article, we have considered a rotating black hole  which is embedded in quintessence universe and hence the black hole metric is constrained with dark energy. A term $\frac{\mathcal{A}_q}{r^{3\omega_q-1}}$ is introduced in the lapse function of the Kerr metric. This term regulates the impact of the back ground exotic fluid into the black hole's attracting nature. Our main motif was to study the nature of fluid flow around this particular type, of black hole. To do this, we consider a pseudo Newtonian approach which will consider the geometric curvature near the black hole but will abort the general relativistic complexity of calculations. Radial momentum balance equation, equation of continuity and the equation of state for the accreting matter are considered and combining them we have obtained two differential equations: One is for radial inward speed and the other is for the speed of sound through the accreting fluid. Azimuthal momentum balance equation turns to be (specific angular momentum) $\lambda$= constant due to the consideration of no viscosity. Vertical momentum balance equation turns to be an adiabatic equation which measures the average disc height in terms of the black hole's attracting force, radial distance from the BH and the speed of sound through the accreting fluid. We have considered a continuous physical flow and to maintain that, the differential equation of radial speed,  if the denominator turns zero at a particular radial distance (coined as critical distance) we have considered the numerator to vanish as well. This helped us to apply L'Hospital's rule to obtain a quadratic of radial velocity slope and thus starting from  the critical point. We may obtain two speed profiles: one for accretion and the other for wind. Upto this part we have applied  general procedures which have been followed in different literature. 

The completely new outcome of the mathematical modeling of the accretion upon the quintessence dominated black hole was realized when we compare the wind speed profiles for different values of the quintessence equation of state $\omega_q$ and the rotational parameter $a$.

We have   divided our analysis in two parts: adiabatic and Chaplygin gas like accreting fluids are considered. For a non-rotating black hole embedded in radiation ($\omega_q=1/3$), we observe that the attraction of the black hole is stronger enough that the critical point be placed far enough to hold the centrifugal outward force due to the rotation. Besides that, the accretion-wind speed at this far shifted critical point is low and the wind speed does not grow much as we move to higher radial distance. As we decrease the value of the equation of state of the background fluid, (i) the critical point is formed nearer to the black hole, (ii) values of critical distance accretion and wind are increased and (iii) wind speed rises much for $x>x_c$, (iv) accretion falls rapidly as $\omega_q$ is decreased. To justify these, results, we infer that the attractive power of the black hole is decreased by the inference of the quintessence and the mare negative pressure of it. This is why we have to move  towards the black hole such that the nearer region's strong attraction can balance the weaker centrifugal force of the nearer region's strong accretion can balance the weaker centrifugal force of the nearer region. The second point signifies that the transonic phase is achieved for high velocity of accretion. Third and fourth observations together imply that the involvement of quintessence weakens accretion and strengthens wind. For rotating black holes, these results are mere prominent. With rotation wind speed for extremely negative quintessence EoS increases rapidly and becomes equal to the speed of light at a finite distance from the black hole. This signifies that the outflow has a speed of light which is preferred to oppose the infall, i.e., the accretion. This reduces the feeding efficiency of the accretion disc. If rotation is higher, the distance where wind becomes equal speedy with light is shifted more towards the black hole. So, the repelling effect of dark energy/quintessence is catalysed by the involvement of the rotation. A point to be noted is that the accreting fluid is of adiabatic type. The quintessence effect is only considered inside the black hole. That too with a regulating factor $\mathcal{A}_q=10^{-10}$. Such a small effect shows this drastic changes. Finally, we have involved dark energy as accreting fluid and we observe the wind to steeply increase after $x>x_c$ to reach the speed of light. We have tried to find out some observational evidences where the accretion is being weaker  with time (i.e., with lowering of the red shift) and wind have become stronger gradually. These results support our theory.

\section*{Acknowledgment}

This research is supported by the project grant of Goverment of West Bengal, Department of Higher Education, Science and Technology and Biotechnology (File no:- $ST/P/S\&T/16G-19/2017$). RB thanks IUCAA for providing Visiting Associateship.

\bibliographystyle{unsrt}

\end{document}